\documentclass{IEEEtran}[10pt,letterpaper,conference]
\usepackage{amsmath}
\usepackage{epsfig}
\title{Modeling queuing dynamics of TCP: a simple model and its empirical validation.}
\author{D. Genin, T. Nakassis\\Email: dgenin@nist.gov, tnakasis@nist.gov}
\begin{document}
\maketitle
\begin{abstract}
Understanding queuing dynamics of TCP is important for correct router buffer sizing as well as for optimizing the performance of the TCP protocol itself. However, modeling of buffer content dynamics under TCP has received relatively little attention given its importance. Commonly used queuing models are based on overly simplistic assumptions about the packet arrival process. As a consequence, there are no quantitatively accurate closed loop TCP models capable of predicting performance even for a single link shared by multiple flows. Our present paper aims to close this gap by proposing a simple TCP queuing model, which is based on experimental observations and validated by extensive packet level simulations.
\end{abstract}
\section{Introduction}
Queuing dynamics of TCP packet traffic has long been a research topic of great interest. Understanding queuing dynamics of TCP is important for correct router buffer sizing as well as for optimizing the performance of the TCP protocol itself. However, modeling of buffer content dynamics under TCP has received relatively little attention given its importance. Commonly used queuing models, on which packet loss models of closed loop TCP models are based, are often ad hoc and are based on overly simplistic assumptions, such as Poisson arrivals of buffer overflow moments and Poisson packet arrivals. As a consequence, there are no quantitatively accurate closed loop models (i.e. mathematical models parametrized solely by the a priori known network parameters, such as router capacity, propagation delay and buffer size) capable of predicting performance even for a single link shared by multiple flows. This is the gap that our present paper aims to fill by proposing a simple TCP queuing model, which is based on experimental observations and validated by extensive packet level simulations.

Before describing the proposed model we give a brief survey of the queuing and packet loss models that have appeared in TCP literature. Mathematical models of TCP require some way of modeling packet loss in order to close the control loop of the TCP congestion avoidance mechanism. Since packet losses occur most often at buffer overflow\footnote{at least in wired networks, which today constitute the bulk of the Internet} it is natural to take packet loss probability as the blocking probability in a single server queue. Thus there is a close connection between packet loss models and queuing models.

There are two distinct approaches to modeling packet loss: stateless and stateful. The former assumes that in equilibrium queue length sampled at the times of packet arrivals behaves as a sequence of i.i.d. random variables. Correctness of this assumption has been experimentally tested for cross-WAN and -Internet TCP flows but it appears to be wrong for LAN flows \cite{AltAvrBar}. On the other hand, the stateful approach assumes that queue length has long range autocorrelation and so the queue state needs to be a part of the whole TCP model. Packet loss probability is then expressed as a function or a random process (depending on the specific model) which depends on the length of the queue.

For the stateless models the queue length distribution is typically assumed to be one of the standard queuing theory models -- M/M/1, M/M/1/B or M/D/1/B. The advantage of these models is the ready availability of explicit analytical expressions for most quantities interest, including the blocking probability. Models based on M/M/1/B queues have been extensively used in publications on fluid approximation, which are numerous, e.g. \cite{KelMauTan},\cite{Vin},\cite{Sri},\cite{JohTan},\cite{DebShaSri},\cite{YinDulSri}, as well as in some papers on the problem of router buffer sizing\cite{EnaGanGoeMcKeoRou},\cite{RaiWis},\cite{RaiTowWis}. It has long been known, however, that these models fail to adequately capture the complex statistical structure of TCP traffic \cite{PaxFlo},\cite{GoeGau},\cite{JiaDov} and produce significantly inaccurate results \cite{GenMar}. We discuss the reason for this failure below. Attempts have been made to modify these models, for example, by adding flow synchronization effects, in order to improve their fidelity but with limited success \cite{BacHon},\cite{BacHon1}.

The stateful queuing models have been especially popular in stochastic TCP modeling literature \cite{BacMcDonJul},\cite{BacMcDonReyJul},\cite{ChaDeVle} but also have appeared in fluid approximation models of TCP with router implementing Random Early Drop\cite{HolMisTowGon},\cite{MisGonTow},\cite{LowPagWangDoy}. These models explicitly include queue length as a variable governed by a coupled (stochastic) differential equation. Packet loss is then modeled as a probability function or a random, usually Poisson, process dependent on the current queue length. While few of the resulting models have been experimentally verified, partially due to their mathematical complexity, they are unlikely to produce accurate results because they are based on overly optimistic assumptions about the uniformity of TCP transmission rate, similar to the M/M/1/B models.

An interesting exception to this rough classification is the model of Dumas, Guillemin and Robert in which the packet loss is assumed to be a Bernoulli process on the RTT\footnote{We use round trip propagation delay and round trip time interchangeably because processing and buffering delays are largely insignificant in the setting we consider.} (round trip time) time scale but has state given by the order of transmission on the sub-RTT scale \cite{DumFabPhi}. Their results, however, give asymptotic formulas for the congestion window size distribution in the limit of zero packet loss for a fixed packet loss probability function, i.e. it is an open loop model, which ignores the problem of relating TCP and network variables to packet loss.

We must also, separately, mention the work of Wischik on router buffer sizing, which attempts to account for the effects of packet burstiness on queue length with a model based on an M/M/1/B queue\cite{Wis}. While burstiness is indeed the crux of the issue, the results of the study have not been experimentally tested, to the best of our knowledge, and the degree of their qualitative and quantitative accuracy remains unclear.

In summary, both the stateless and the stateful queuing models are based on fundamentally flawed assumptions about the uniformity of the TCP packet emission process. This observation has been made by a number of authors \cite{HuaLiuGonTow},\cite{GoeGau},\cite{GenMar}. Most explicitly the issue has been described by Baccelli and Hong in \cite{BacHon1}. There the authors describe a large scale fluid approximation TCP/IP network simulator based on the M/M/1/B queuing model with a modification accounting for increased packet loss due to flow synchronization. The simulator performed well when the access link speed was not too high, however, when it became larger than a certain threshold the simulator significantly overestimated the throughput because ``with high speed local links, packets are very likely to be concentrated at the beginning of RTTs. Such a packet concentration creates losses even if the input rate averaged over one RTT is much smaller than the capacity of the shared resource."\cite{BacHon1}

The origin of the burstiness observed by Baccelli and Hong, and others is easily traced to the operation of the sliding window algorithm used by TCP. This algorithm releases packets for transmission only in response to acknowledgments received for previously sent packets. The result of this behavior is that packets flow back and forth in formation, creating bursts followed by silences, during which the algorithm awaits the acknowledgments of the transmitted packets. Instead of sliding smoothly along the queue of packets ready to be transmitted, the sliding window instead moves in fits and starts. We note, that the sliding window algorithm is implemented in all current TCP variants. Therefore, we can be sure that this transmission burstiness affects not only the long standard TCP-Reno but every implementation of TCP in existence. The only modification that could change this would be transmission pacing, but it may have issues of its own\cite{Wis},\cite{Agg}.

The present paper is meant to fill a certain gap in the program we envision as leading to high-fidelity TCP networking models. It has long been assumed that mean value models, such as fluid approximation, can give an accurate estimate of the steady state, and even dynamics of, TCP performance. Recent research, however, indicates that statistics of the TCP congestion window size and of the transmission process contribute significantly even when the number of flows is very large, and that more sophisticated models are necessary if quantitatively, as well as qualitatively, accurate results are desired. The ingredients necessary to achieve this goal, in our view, are
\begin{enumerate}
\item[i)] an accurate model for the congestion window size distribution as a function of the network parameters and packet loss;
\item[ii)] an accurate queuing model that can be used to deduce the packet loss probability in terms of the congestion window size distribution and parameters of the network.
\end{enumerate}
These correspond to the two halves of the TCP congestion avoidance feedback loop. Significant progress has been made by a number of authors in tackling the first item. Closed, though, still rather mathematically complex, expressions have been obtained for the stationary distribution of the congestion windows sizes of TCP flows under the restriction of a constant RTT \cite{BacMcDonJul},\cite{DumFabPhi},\cite{ChaDeVle}. Our present work is aimed at providing a partial answer to the second problem.

The on-off fluid source model we propose here is not new, although some adaptation to the specifics of TCP was necessary. We describe the model in detail in Section \ref{model}, but we highlight here the observations about the equilibrium TCP queuing behavior that lead to it:
\begin{enumerate}
\item[1)] in equilibrium TCP sources can be treated as statistically independent\footnote{Synchronization between flows does occur for certain network parameter combinations in our study but is atypical (see Figure \ref{sync}).},
\item[2)] in equilibrium a TCP source can be treated as a stationary random fluid on-off source.
\end{enumerate}
The first observation is perhaps not very surprising given that the flows are coupled only by packet loss at the queue, which at equilibrium typically behaves like a stationary random process. Observation (2), however, is somewhat less obvious, as it says that the exact congestion window trajectories of individual flows do not matter to the router queue length distribution. That is, from the point of view of the buffer, a collection of ordinary TCP sources and ones whose congestion windows change randomly, but follow the same distribution as the former, are equivalent\footnote{provided the number of sources is sufficiently large}.

These observations permit construction of a simple yet accurate TCP queuing model. The basic structure of the model follows that of Kosten \cite{Kos} and Anik, Mitra and Sondhi \cite{AniMitSon}. The main difference with the former is that the ``on" and ``off" periods are not exponentially distributed, which makes exact solution for the stationary queue length distribution unlikely. Still, there is hope that some analytical headway can be made with more general distributions based on the work of Palmowski and Rolski\cite{PalRol}. The proposed model also has the advantage of providing a unified framework for treating large as well as small buffer regimes, which so far have usually been treated as distinct asymptotic regimes requiring separate calculations.

We validate the Kosten-Anik-Mitra-Sondhi (KAMS) model by comparing the stationary queue length distribution obtained from \emph{ns2} simulations with that obtained from numerical simulations of the model for a large number of network parameter combinations. Since we measure the distance between whole queue length probability distributions, which are signatures of the underlying queuing processes, the near perfect observed match between distributions is an indication of the identity of the queuing processes that produced them.

This work is a refinement of our previous paper \cite{GenMar}, where we first applied KAMS to TCP queuing and provided experimental evidence in its support. Although our work is independent, we have recently learned that several other authors have also arrived at related models, in our view, strengthening the case for its correctness and utility \cite{HuaLiuGonTow},\cite{GoeGau}.

The rest of the paper is arranged as follows. In Section \ref{model} we layout the details of the KAMS model and its adaptation to TCP. Section \ref{methodology} describes details of the validation approach. Results of validation are presented in Section \ref{results}. Section \ref{conclusion} summarizes our findings and discusses directions for future work.

\section{KAMS TCP Queuing Model}
\label{model}

The model we propose for TCP queuing is derived from the multi-input buffer queuing model originally studied by Kosten \cite{Kos} and later by Anik, Mitra and Sondhi \cite{AniMitSon} (and many others since). The main advantage of KAMS is that it is able to incorporate TCP burstiness, essential to accurate queue modeling, with a minimum of mathematical complexity.

The basic, in the topological sense, KAMS model is a network with a single server and a fixed number of inputs. This admittedly simple setup is a fundamental building block in and stepping stone to the study of more complicated TCP networks. Even this topologically simple model can be practically useful for modeling bottleneck routers in the broadband edge network. 

The queue input process in KAMS is modeled as a superposition of some number, $N$, of statistically independent random on-off fluid sources. Each source has two states --- "on", when it is transmitting, and "off", when it is not. In the on state, every source emits data with some constant rate $\nu$. 

This two level fluid source model is a natural fit for approximating bursty TCP transmission behavior. The on periods correspond to packet-burst transmissions, when the TCP source transmits the packets one after the other in quick succession, and the off periods to silences as the source awaits acknowledgment of the transmitted packets. 

Data streams from all sources merge in the server queue so that when $k$ sources are active the total data inflow rate is $k\nu$. The server processes incoming data at a constant rate $C$. When $k\nu>C$ the arrived but unprocessed data is stored in a buffer of size $B$. Any data arriving after the buffer becomes full is discarded.

\begin{figure}[ht]
\epsfig{file={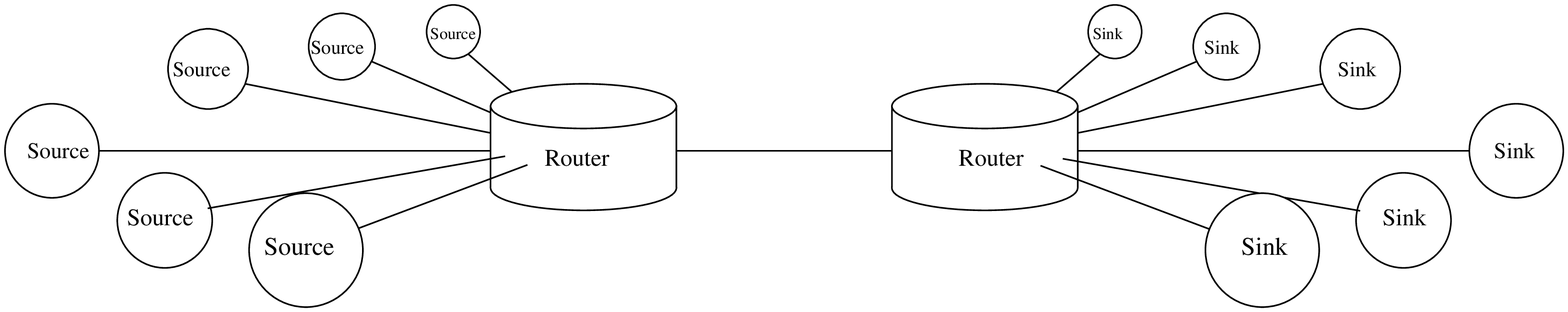},width=8cm}
\caption{Basic network topology.}
\label{network}
\end{figure}

The equivalent computer network looks like a double fan (Fig. \ref{network}). The TCP sources on the left transmit data to their respective receivers on the right over the bottleneck link in the middle. To make this basic architecture as simple as possible we will also assume that round trip propagation delays between the source-receiver pairs are roughly equal. This is the archetypal bottleneck network. 

Most of the KAMS parameters match up naturally with this model bottle network's parameters, so $C$ is the bottleneck router capacity, $B$ is the router buffer size and $N$ the number of TCP sources. The three remaining parameters --- $\nu$, and the random processes determining the on and off period durations require specific tuning to match KAMS behavior to the model TCP network.

In the original KAMS model $\nu$ is a free parameter, permitted to vary from source to source. For a TCP flow the rate of transmission is determined by the parameters of the network, provided a source's rate is not itself the limiting factor. The timing of packet emissions is determined by the TCP sliding window algorithm. Since a new packet is sent only when an acknowledgment for the previous packet is received, the speed of the access link, which is usually the slowest segment of the network, determines the interval between consecutive packets emissions. We will assume that the maximum speed of a TCP source is not less than the speed of its access link. Thus $\nu$ is equal to the speed of the access link. In practice, this assumption is vacuous, since the speed of the access link can be assumed to be equal to the speed of the source if the source is slower.

Duration of a given on period is determined by the size of the packet burst. Further, the number of packets in a burst, due to idiosyncrasies of the sliding window algorithm, is equal to the size of the sliding window, which we will assume is equal to the congestion window. Indeed, analysis of \emph{ns2} packet traces confirms that most packets are transmitted in bursts equal in size to the congestion window \cite{GenMar}. Moreover, packets in a burst are transmitted back to back at the speed of the slowest router in the path, i.e. $\nu$. Thus a duration of a given on period is simply the congestion window size at the time of its initiation divided by $\nu$.

It turns out that due to the multiplexing at the router it is not necessary to have the on periods of a given source follow the additive increase multiplicative decrease(AIMD) of the TCP congestion avoidance algorithm. Provided the sources activate independently of each other, the on periods of a given source can be modeled as a sequence of i.i.d. random variables with the same distribution as a deterministic AIMD source. From the point of view of the router these two input processes appear to be indistinguishable. The problem of determining the on period duration thus shifts to identifying the stationary congestion window size distribution.

Several papers have been published describing the stationary congestion window size distribution for a single source as well as for a large number of sources sharing a bottleneck link. The resulting mathematical expressions are, unfortunately, highly complex, making it difficult to check their identity or difference. Their complexity also makes it very difficult to implement realizations of the corresponding random variables. In practice, a good approximation is often sufficient for obtaining very accurate results. One characteristic that all theoretically computed distributions share is the Gaussian tail. We, therefore, postulate that the congestion window size distribution is approximated by a truncated normal distribution, restricted to the range $[0,\infty)$. Lacking a ready formula relating the parameters of this distribution to the network parameters of the model, we estimated them instead from experimental data. Details of the estimation procedure are explained in Section \ref{methodology}. We emphasize that this is not a shortcoming of the proposed model, since it is only meant to model the steady state queuing statistics and \emph{not} the TCP congestion avoidance algorithm. 

Finally, the off period is the silence between consecutive bursts of packets. Since at most a congestion window's worth of packets can be sent in a single round trip and since this is exactly the burst size observed it follows that the duration of an off period is equal to RTT minus the duration of the last burst. However, because we will be concerned with high speed routers the burst duration will be negligibly small compared to the RTT. Since the same applies to the buffering delay, the off periods can be assumed to be equal to the round trip propagation delay, identical and constant for all sources.

\section{Methodology}
\label{methodology}
To validate the constructed model we compared the queue length distributions obtained from numerical simulations of the KAMS model with the same distributions computed from \emph{ns2} packet level simulations for a range of network parameter combinations. Below we describe the parameter values used in the validation study and the rational behind the specific choices.

The number of flows $N$ in the validation study was set to 1000. Firstly, this is a large number of flows that might realistically be observed at a router at the periphery of the Internet, where most bottlenecks lie. Secondly, at $N=1000$ the model is close enough to the asymptotic limit that increasing the number of flows further does not significantly affect the queue length distribution. Router capacity was fixed at $C=1$ Gbps. Access link speed was set at 100 Mbps, which gives $\nu=100$ Mbps.

The remaining scalar parameters --- router buffer size and round trip propagation delay --- were varied to determine how accurately the KAMS model tracks packet level simulations. These parameters are also known to have the strongest influence on TCP behavior and so we were most interested in the fidelity of the KAMS model with respect to them.

Router buffer size was varied in steps of 50 pkts from 50 pkts to 300 pkts. Preliminary experiments with \emph{ns2} suggested that increasing buffer size beyond 300 pkts does not significantly affect packet loss and so 300 pkts was chosen as an upper limit on the buffer size. Conversely, for buffer sizes below 50 pkts packet loss increases dramatically, which makes it a reasonable lower bound.

The round trip propagation delay was varied between 50 ms and 300 ms in increments of 50 ms. This range of propagation delays corresponds approximately to wired networks varying in diameter from a LAN to the global Internet.

Finally, to fix the on period duration distribution it was necessary to determine the parameters of the truncated normal distribution, which we postulated approximates the congestion window size distribution. Analyzing \emph{ns2} experimental data we found that the congestion window size distribution is, indeed, very well approximated by a truncated normal distribution. The mean of the distribution was take to be the mode of the empirical distribution, computed as the average of the four most likely values. The variance was then estimated by performing a least squares fit of the truncated normal distribution with the predetermined mean. The fit was performed on the logarithm of the data, which gives a greater weight to the accurate fitting of the tail of the distribution. In this way the mean and variance of the congestion window size distribution were computed for each of the 36 pairs of propagation delay and buffer size to be used for running the KAMS model simulation with corresponding parameters.

To summarize, the KAMS model was tested with $N=1000$ flows, $C=1$Gbps, $\nu=100$Mbps, router buffer size ranging from 50 pkts to 300 pkts in steps of 50 pkts and RTT ranging from 50 ms to 300 ms in steps of 50ms, giving 36 test points.

Validation was performed by running simulations of \emph{ns2} and KAMS for each of the 36 test points for 600 simulated seconds. In \emph{ns2} simulations queue length was sampled once per round trip. The KAMS simulation sampled queue length at the moments when sources changed state. The cumulative queue length distributions were computed by binning the queue length data into consecutive integer bins. Only data from the final 80\% of the simulated time interval was used to erase the effects of the transient phase.

\section{Results}
\label{results}
\begin{figure}[ht]
\epsfig{file=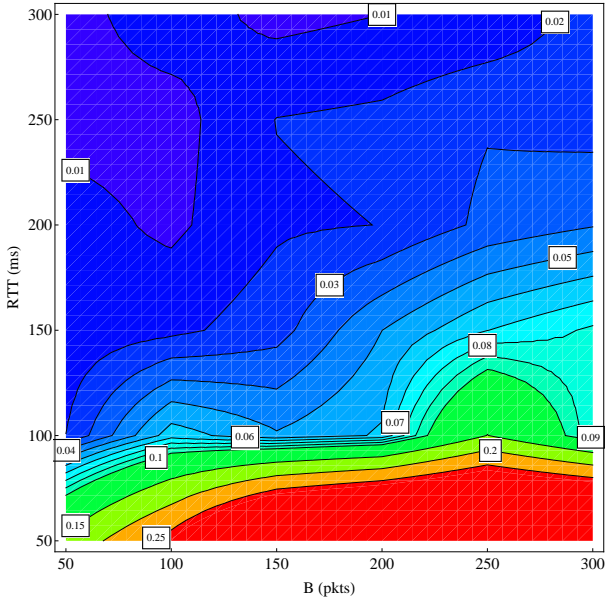,width=8cm}
\caption{Contour plot of NRMSE in KAMS cumulative queue length distribution.}
\label{NRMSE}
\end{figure}

Results of the comparison between KAMS and \emph{ns2} simulations are summarized in Figure \ref{NRMSE}. The figure contains the contour plot of normalized root mean square error (NRMSE) in KAMS cumulative queue length distribution relative to \emph{ns2} over queue lengths greater than 5 packets.
We chose to drop the distribution values for the queue lengths between 0 and 5 because they are of little practical importance but contribute disproportionately to the error. Including the discarded values does increase NRMSE by about 5\%. The reason for the disproportionately high contribution is that near the buffer boundaries, i.e. near 0 and $B$, the accuracy of the fluid approximation breaks down. This is because KAMS stationary probability distribution becomes singular at the buffer boundaries, giving rise to jump discontinuities in the cumulative distribution. On the other hand, the \emph{ns2} cumulative queue length distribution, while rising sharply near the boundaries, remains continuous. The result is a very large relative error in the boundary regions. This becomes a problem especially when attempting to estimate the buffer overflow probability.

As can be seen from Figure \ref{NRMSE} the KAMS model exhibits a remarkably high accuracy --- better than 5\% NRMSE --- for all but two parameter combinations with round trip propagation delay $\geq$100 ms. (Experiments with $\nu=C$ produce similar results which we do not present here for the sake of brevity.)

\begin{figure}[ht]
\epsfig{file=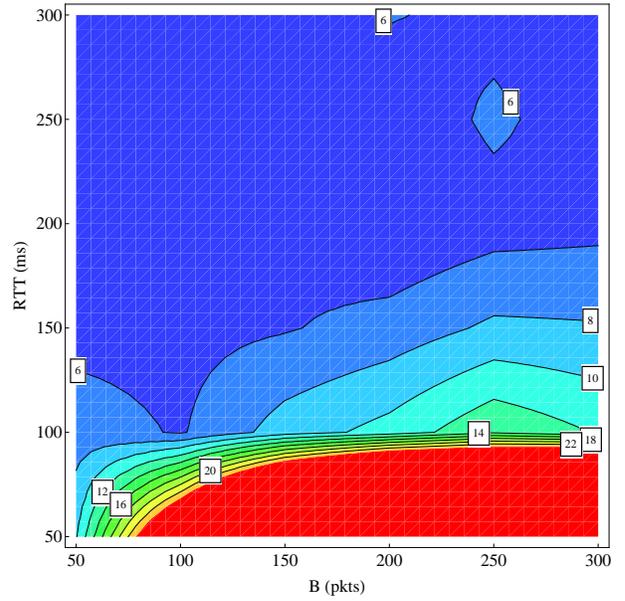,width=8cm}
\caption{Contour plot of the degree of synchronization of aggregate packet loss. Higher values correspond to more synchronization.}
\label{sync}
\end{figure}

An interesting observation emerges from the analysis of failure of KAMS in the region RTT$<100$ ms. \emph{Ns2} packet loss time series data indicate that for RTT=50 ms there is significant packet loss synchronization between flows. Figure \ref{sync} attempts to capture the degree of synchronization of aggregate packet loss with one number. The function, whose contours appear in the figure, is a simple measure of ``spikiness" of the Fourier transform of the aggregate packet loss time series. We defined it as
\begin{equation}
\frac{\max_{i>0} |Re(\omega_i)|}{\frac{1}{M}\sum_{i>0}|Re(\omega_i)|}
\end{equation}
where $\omega_i$ is the $i$th Fourier coefficient of the aggregate packet loss time series and $M$ is the length of the data series. The largest values, occurring for RTT=50 ms, correspond to spectral densities with pronounced peaks arising from strong periodicity in the packet loss time series. This agrees with observations made in \cite{AltAvrCha}, where authors conclude that for long RTTs packet losses are well approximated by an i.i.d. random process, whereas for short RTTs there is a significant long-range time correlation between packet losses.

Counterintuitively, KAMS errs on the side of higher queue lengths and higher packet losses, which is the opposite of what we expected to see when buffer overflow events become synchronized. Synchronization is still a poorly understood phenomenon and why it occurs for short RTTs, when the lag in the congestion avoidance control loop is small, rather than for long RTTs is a topic for future research.

\begin{figure}[ht]
\epsfig{file=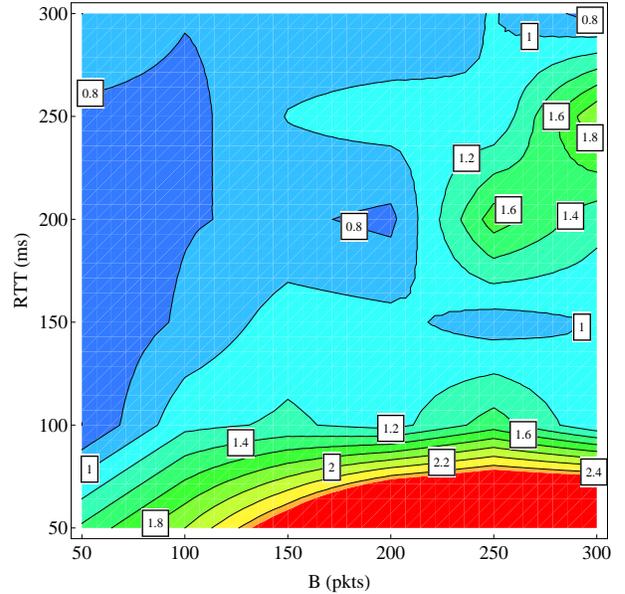,width=8cm}
\caption{Multiplicative error in corrected full buffer probability of KAMS.}
\label{full_buffer_error}
\end{figure}

Ultimately, the point of TCP queuing models is to provide a formula approximating the buffer overflow probability and, hopefully, observed packet loss probability. Hence, we consider next the accuracy of KAMS full buffer probability. As pointed out above, KAMS performs poorly near the buffer boundaries. The relative error in the full buffer probability is between 600 and 1400\%. However, the error is nearly constant across most of the test parameter range, allowing us to introduce a scaling correction factor. Figure \ref{full_buffer_error} show the contour plot of the multiplicative error in KAMS full buffer probability data after rescaling by the correction factor. The correction factor was chosen equal to the average of the multiplicative error for RTT$\ge 100$ ms, to avoid skewing the mean by the abnormally high values from the parameter combinations exhibiting synchronization. As can be seen from the plot, for most test points the multiplicative error drops below 20\%. Since in the TCP throughput formula packet loss probability appears under the square root, the error in estimated throughput would be smaller still, on the order of 10\%.

We note that the \emph{ns2} observed full buffer probability does not equal the observed mean packet loss rate, although, in the absence of synchronization, they are roughly proportional with a coefficient of about 0.6. Understanding of the relationship between full buffer probability and packet loss requires further research.

We also note that the KAMS model is highly sensitive to the shapes of the on and and off period distributions. Initially, we tried using exponential distributions for both since a closed formula already exists for the stationary queue length distribution in this setting. The results were considerably worse with NRMSE running as high as 20\% for many test points. Even using a half-normal distribution (i.e. a truncated normal distribution with mean 0) produced results that were substantially worse, even though the means of the truncated normals used above were usually in the low teens. Switching an exponential to constant off period distribution had a less dramatic but still noticeable.

\section{Conclusion}
\label{conclusion}
Overall, when the assumptions of the model are satisfied, KAMS can be said to approximate the stationary queue length distribution exceptionally well. This, in our opinion, is a strong indication that KAMS correctly represents the TCP queuing process. For well understood reasons, in absolute terms KAMS is less accurate at the upper and lower buffer limits. However, the KAMS full buffer probability is roughly proportional to the observed one with a universal (across network parameters) constant coefficient.

The ultimate goal of this work is to facilitate the creation of numerically accurate closed form mathematical models of TCP networks. In light of the presented work, two more problems need to be resolved to achieve this final goal: first, a closed form mathematical expression approximating buffer overflow probability in the KAMS model, and, second, a closed form approximation expressing the relationship between mean and variance of the truncated normal distribution, representing the stationary congestion window size distribution, and packet loss probability. Headway, has already been made in both of these directions by Palmowski and Rolski in establishing results on the statistics of the stationary KAMS distribution for general on-off period distributions\cite{PalRol}, and by Bacelli, Dumas, Chaintreau and others in establishing the relationship between the packet loss process and stationary congestion window size distribution \cite{BacMcDonJul},\cite{DumFabPhi},\cite{ChaDeVle}. The final synthesis of these elements into a highly accurate model of TCP networking is a topic for future research.

\bibliographystyle{plain}
\bibliography{fluid_model_taxonomy}
\end{document}